\documentclass[showpacs,amsmath,preprint]{revtex4-1}

\def\be{\begin{equation}}
\def\ee{\end{equation}}
\def\ba{\begin{array}}
\def\ea{\end{array}}

\def\dps{\displaystyle}

\def\qed{\leavevmode\unskip\penalty9999 \hbox{}\nobreak\hfill
     \quad\hbox{\leavevmode  \hbox to.77778em{%
               \hfil\vrule   \vbox to.675em%
               {\hrule width.6em\vfil\hrule}\vrule\hfil}}
     \par\vskip3pt}
\usepackage{graphicx,epsfig,dsfont,amssymb,amsmath,amsthm,amsfonts,amsbsy,mathrsfs}
\newtheorem{theorem}{Theorem}

\newtheorem{corollary}{Corollary}

\begin{document}

\title{Detection of Ideal Resource for Multiqubit Teleportation}

\author{Ming-Jing Zhao$^{1}$}
\author{Bin Chen$^{2}$}
\author{Shao-Ming Fei$^{2,3}$}

\affiliation{
$^1$Department of Mathematics, School of Science, Beijing Information Science and Technology University, 100192, Beijing,
China\\
$^2$School of Mathematical
Sciences, Capital Normal University, Beijing 100048, China\\
$^3$ Max-Planck-Institute for Mathematics in the
Sciences, 04103 Leipzig, Germany\\}

\begin{abstract}
We give a sufficient condition in detecting entanglement resource for perfect multiqubit teleportation.
The criterion involves only local measurements on some complementary observables and can be experimentally
implemented. It is also a necessary condition for fully separability of multiqubit states.
Moreover, by proving the optimality of teleportation witnesses, we solve the open problem in [Phys. Rev. A {\bf 86}, 032315 (2012)].

\end{abstract}

\maketitle

\section{Introduction}
Quantum teleportation, employing classical communication and shared
resource of entanglement, allows to transmit an unknown quantum
state from a sender to a receiver that are spatially separated. The first teleportation protocol was proposed by
using Einstein-Podolsky-Rosen pair by Bennett {\it et.
al.} \cite{C. H. Bennett}. Then the three-qubit GHZ state and a
class of W states are revealed to be the ideal resource for faithful
teleportation of one-qubit state \cite{A. Karlsson,P. Agrawal}. For
two-qubit state teleportation, the tensor product of two Bell states
\cite{G. Rigolin} and the genuine four-qubit entangled state \cite{Y.
Yeo} are showed to have the ability of faithful teleportation.
For three-qubit state, Refs. \cite{S. Choudhury,X. Yin} have investigated its teleportation
in terms of a genuine entangled six-qubit state. Some entangled $2n$-qubit states have been presented for
$n$-qubit state teleportation \cite{S. Muralidharan2010, M. J. Zhao12jpa,M. Y. Wang}.

One important problem related to quantum teleportation is how to know if a general quantum state is the ideal resource
for faithful teleportation, or is useful for teleportation with fidelity better than classical states.
For bipartite case, it has been shown that the quantum states which are not useful for quantum teleportation compose a compact convex set
and a teleportation witness has been presented for the first time in Ref. \cite{N. Ganguly}.
Then a complete set of teleportation witness to detect all the ideal resource for teleportation is constructed in
Ref. \cite{M. J. Zhao12pra}. In Ref. \cite{S. Adhikari}, a systematic method to construct teleportation witness from
entanglement witness is provided. The properties of teleportation witness are further studied in Ref. \cite{A. Kumar}.
For multipartite case, Ref. \cite{P. X. Chen} provides the necessary and sufficient condition that the genuine $2n$-qubit
entanglement channels must satisfy for teleporting an arbitrary $n$-qubit state,
and Refs. \cite{C. Y. Cheung,D. M. Liu} analyze the criterion of multiqubit states for $n$-qubit teleportation.

In this paper, we first study the multiqubit teleportation and propose a linear operator to detect ideal resource for multiqubit teleportation.
This operator is expressed by the local complementary observables. It gives a sufficient condition for the detection of all
ideal resource for multiqubit teleportation. This operator can also be used to detect multiqubit entanglement and serves as a necessary condition for fully separability.
Then we analyze teleportation witness for bipartite high dimensional case. Moreover, the problem left in Ref. \cite{S. Adhikari} is solved
by explaining the optimality of teleportation witness.

\section{Detection of Ideal Resource for Multiqubit Teleportation}

We first consider $n$-qubit systems. Let $\vec{a}_1$, $\vec{a}_2$ and $\vec{a}_3$ be three dimensional unit vectors.
We say that three observables $X_k=\vec{a}_k\cdot\vec{\sigma}$, $k=1,2,3$, are complementary if they satisfy
$X_1X_2X_3=-{\rm i}I_2$, where ${\rm i}=\sqrt{-1}$, $I_2$ is the $2\times 2$ identity matrix,
$\vec{\sigma}=(\sigma_1, \sigma_2, \sigma_3)$, $\sigma_1=|0\rangle\langle1|+|1\rangle\langle0|$,
$\sigma_2={\rm i}(|0\rangle\langle1|-|1\rangle\langle0|)$, $\sigma_3=|0\rangle\langle0|-|1\rangle\langle1|$ are Pauli matrices,
which means that the orientation of the basis formed by three real vectors
$\{\vec{a}_k\}_{k=1}^3$ is right-handed, with the same orientation as that of three Pauli matrices.
Let $\{X_k^{A_i}\}_{k=1}^3$, $i=1,\cdots,n$, be complementary observables acting on $A_i$.
We define the operator,
\begin{equation}
\begin{array}{rcl}\label{th eq}
&&\Gamma\equiv\frac{1}{2^{2n}}\{1+
\sum_i [X_1^{A_i} \otimes \sigma_1^{B_i}-X_2^{A_i} \otimes \sigma_2^{B_i}+X_3^{A_i} \otimes \sigma_3^{B_i}]\\&&\dps
+ \sum_{i_1,i_2}  [\sum_{j_1,j_2=1,3}X_{j_1}^{A_{i_1}} \otimes X_{j_2}^{A_{i_2}} \otimes \sigma_{j_1}^{B_{i_1}}
\otimes \sigma_{j_2}^{B_{i_2}}-\sum_{j=1,3}X_j^{A_{i_1}} \otimes X_2^{A_{i_2}} \otimes \sigma_j^{B_{i_1}}\otimes \sigma_2^{B_{i_2}}\\&&\dps
+ X_2^{A_{i_1}} \otimes X_2^{A_{i_2}} \otimes \sigma_2^{B_{i_1}}\otimes
\sigma_2^{B_{i_2}}]\\&&\dps
+\cdots
+\sum_{i_1,\cdots, i_s} [\sum_{j_1,\cdots,j_s=1,3}X_{j_1}^{A_{i_1}} \otimes\cdots\otimes X_{j_s}^{A_{i_s}} \otimes
\sigma_{j_1}^{B_{i_1}}\otimes\cdots\otimes \sigma_{j_s}^{B_{i_s}}\\&&\dps
-\sum_{j_1,\cdots,j_{s-1}=1,3}X_{j_1}^{A_{i_1}} \otimes \cdots \otimes X_{j_{s-1}}^{A_{i_{s-1}}}\otimes X_2^{A_{i_s}} \otimes \sigma_{j_1}^{B_{i_1}} \otimes \cdots \otimes\sigma_{j_{s-1}}^{B_{i_{s-1}}}\otimes \sigma_2^{B_{i_s}}+\cdots\\&&\dps
+(-1)^s X_2^{A_{i_1}} \otimes\cdots\otimes X_2^{A_{i_s}}
\otimes \sigma_2^{B_{i_1}}\otimes\cdots\otimes \sigma_2^{B_{i_s}}
]\\&&\dps
+\cdots
+[\sum_{j_1,\cdots, j_n=1,3}X_{j_1}^{A_1} \otimes\cdots\otimes X_{j_n}^{A_n} \otimes \sigma_{j_1}^{B_1}\otimes\cdots\otimes
\sigma_{j_n}^{B_n}\\&&\dps
-\sum_{i_1,\cdots, i_n}\sum_{j_1,\cdots, j_{n-1}=1,3}X_{j_1}^{A_{i_1}} \otimes \cdots \otimes X_{j_{n-1}}^{A_{i_{n-1}}} \otimes X_2^{A_{i_n}} \otimes \sigma_{j_1}^{B_{i_1}} \otimes \cdots \otimes\sigma_{j_{n-1}}^{B_{i_{n-1}}} \otimes \sigma_2^{B_{i_n}}\\&&\dps
+\cdots+(-1)^n X_2^{A_1} \otimes\cdots\otimes X_2^{A_n} \otimes \sigma_2^{B_1}\otimes\cdots\otimes \sigma_2^{B_n}]\}.
\end{array}
\end{equation}
The operator $\Gamma$ can be used to detect the ideal resources for $n$-qubit teleportation.

\begin{theorem}\label{th1}
Any $2n$-qubit pure state $\rho=|\phi\rangle_{A_1\cdots A_n,B_1\cdots B_n}\langle\phi|$, with qubits $A_i$ in $A$ part and qubits $B_i$ in $B$ part respectively,
can be used to teleport an arbitrary $n$-qubit state faithfully if there exist complementary observables $\{X_k^{A_i}\}_{k=1}^3$ acting on $A_i$ such that
$\langle \Gamma\rangle_{\rho}=1$.
\end{theorem}

{\sf [Proof].} First, as is well known, the tensor product of Bell states,
\begin{eqnarray*}
|\phi^+\rangle_{A_1\cdots A_n,B_1\cdots B_n}
=\frac{1}{\sqrt{2^n}}\sum_{i_1,\cdots, i_n=0,1} |i_1,\cdots, i_n\rangle_{A_1\cdots A_n} |i_1,\cdots, i_n\rangle_{B_1\cdots B_n},
\end{eqnarray*}
is an ideal resource for multiqubit teleportation. Let us expand the operator $|\phi^+\rangle_{A_1\cdots A_n,B_1\cdots B_n}\langle\phi^+|$
in terms of the Pauli operators. Note that the expectation values of $\sigma_k$ with respect to $|\phi^+\rangle_{A_1\cdots A_n,B_1\cdots B_n}\langle\phi^+|$
are all zero expect for the case that there are even number of $\sigma_k$s, with one half acting on some qubits of $A$ part and the other half acting
on the corresponding qubits of $B$ part, namely,
\begin{eqnarray*}
Tr(\sigma_{k}^{A_{i_1}}\otimes\cdots\otimes\sigma_{k}^{A_{i_s}}\otimes \sigma_{k}^{B_{i_1}}
\otimes\cdots\otimes\sigma_{k}^{B_{i_s}}|\phi^+\rangle_{A_1\cdots A_n,B_1\cdots B_n}\langle\phi^+|)=1,
\end{eqnarray*}
$k=1,3$,
and
\begin{eqnarray*}
Tr(\sigma_{2}^{A_{i_1}}\otimes\cdots\otimes\sigma_{2}^{A_{i_s}}\otimes \sigma_{2}^{B_{i_1}}
\otimes\cdots\otimes\sigma_{2}^{B_{i_s}}|\phi^+\rangle_{A_1\cdots A_n,B_1\cdots B_n}\langle\phi^+|)=(-1)^{s},
\end{eqnarray*}
$s=1,2,\cdots,n$. For the case of different Pauli operators acting on the same $t$-th qubit in $A$ and $B$ parts, consider
$\sigma_k$ acts on the qubit $A_t$ and $\sigma_{k^\prime}$ acts on the qubit $B_t$. Then the expectation value

\begin{eqnarray*}
Tr(\sigma_{i_1}^{A_{1}}\otimes\cdots\otimes \sigma_k^{A_t}\otimes \cdots\otimes\sigma_{i_n}^{A_{n}}\otimes \sigma_{j_1}^{B_{1}}
\otimes\cdots\otimes \sigma_{k^\prime}^{B_t}\otimes \cdots\otimes\sigma_{j_n}^{B_{n}}|\phi^+\rangle_{A_1\cdots A_n,B_1\cdots B_n}\langle\phi^+|)=0,
\end{eqnarray*}

due to $\sum_{i_t,j_t=0,1} \langle i_t |\sigma_k|j_t\rangle \langle i_t|\sigma_{k^\prime} |j_t\rangle=0$ for $k\neq k^\prime$, $k, k^\prime=0,1,2,3$,
and $\sigma_0$ is the identity operator. From the above analysis, we have

\begin{equation}\label{phi+ expansion}
\begin{array}{rcl}
&&\dps|\phi^+\rangle_{A_1\cdots A_n,B_1\cdots B_n}\langle\phi^+|=\frac{1}{2^{2n}}\{1+
\sum_i [\sigma_1^{A_i} \otimes \sigma_1^{B_i}-\sigma_2^{A_i} \otimes \sigma_2^{B_i}+\sigma_3^{A_i} \otimes \sigma_3^{B_i}]\\&&\dps
+ \sum_{i_1,i_2}  [\sum_{j_1,j_2=1,3}\sigma_{j_1}^{A_{i_1}} \otimes \sigma_{j_2}^{A_{i_2}} \otimes \sigma_{j_1}^{B_{i_1}}
\otimes \sigma_{j_2}^{B_{i_2}}-\sum_{j=1,3}\sigma_j^{A_{i_1}} \otimes \sigma_2^{A_{i_2}} \otimes \sigma_j^{B_{i_1}}\otimes \sigma_2^{B_{i_2}}\\&&\dps+ \sigma_2^{A_{i_1}} \otimes \sigma_2^{A_{i_2}} \otimes \sigma_2^{B_{i_1}}\otimes
\sigma_2^{B_{i_2}}]\\&&\dps
+\cdots
+\sum_{i_1,\cdots, i_s} [\sum_{j_1,\cdots,j_s=1,3}\sigma_{j_1}^{A_{i_1}} \otimes\cdots\otimes \sigma_{j_s}^{A_{i_s}} \otimes
\sigma_{j_1}^{B_{i_1}}\otimes\cdots\otimes \sigma_{j_s}^{B_{i_s}}\\&&\dps
-\sum_{j_1,\cdots,j_{s-1}=1,3}\sigma_{j_1}^{A_{i_1}} \otimes \cdots \otimes\sigma_{j_{s-1}}^{A_{i_{s-1}}}\otimes \sigma_2^{A_{i_s}} \otimes \sigma_{j_1}^{B_{i_1}} \otimes \cdots \otimes\sigma_{j_{s-1}}^{B_{i_{s-1}}}\otimes \sigma_2^{B_{i_s}}+\cdots\\&&\dps
+(-1)^s \sigma_2^{A_{i_1}} \otimes\cdots\otimes \sigma_2^{A_{i_s}}
\otimes \sigma_2^{B_{i_1}}\otimes\cdots\otimes \sigma_2^{B_{i_s}}
]\\&&\dps
+\cdots
+[\sum_{j_1,\cdots, j_n=1,3}\sigma_{j_1}^{A_1} \otimes\cdots\otimes \sigma_{j_n}^{A_n} \otimes \sigma_{j_1}^{B_1}\otimes\cdots\otimes
\sigma_{j_n}^{B_n}\\&&\dps
-\sum_{i_1,\cdots, i_n}\sum_{j_1,\cdots, j_{n-1}=1,3}\sigma_{j_1}^{A_{i_1}} \otimes \cdots \otimes\sigma_{j_{n-1}}^{A_{i_{n-1}}} \otimes \sigma_2^{A_{i_n}} \otimes \sigma_{j_1}^{B_{i_1}} \otimes \cdots \otimes\sigma_{j_{n-1}}^{B_{i_{n-1}}} \otimes \sigma_2^{B_{i_n}}\\&&\dps+\cdots+(-1)^n \sigma_2^{A_1} \otimes\cdots\otimes \sigma_2^{A_n} \otimes \sigma_2^{B_1}\otimes\cdots\otimes \sigma_2^{B_n}]\}.
\end{array}
\end{equation}

Denote $(U^{ A_i}){^\dagger}\sigma_k U^{A_i}=X_k^{A_i}$. Then $\{X_k^{A_i}\}_{k=1}^3$ are complementary observables
satisfying $X_1^{A_i}X_2^{A_i}X_3^{A_i}=-{\rm i}I_2$, $i=1,2,\cdots,n$. The unitary operators and the complementary observables
can be mutually determined uniquely. If $\langle \Gamma\rangle_{\rho}=1$ for some local complementary observables, then there exist
unitary operators $U^{A_1}$, $\cdots$, $U^{A_n}$ such that
$$
U^{A_1}\otimes\cdots\otimes U^{A_n}|\phi\rangle_{A_1\cdots A_n,B_1\cdots B_n}\langle\phi|U^{A_1\dagger}\otimes\cdots\otimes U^{A_n\dagger}
$$
is of the form (\ref{phi+ expansion}). Therefore, $\rho$ is local unitarily  equivalent to $|\phi^+\rangle_{A_1\cdots A_n,B_1\cdots B_n}\langle\phi^+|$ and is an
ideal resource for multiqubit teleportation. \qed

Theorem \ref{th1} provides a method to detect ideal resource for multiqubit teleportation. All quantum states $|\phi\rangle_{A_1\cdots A_n,B_1\cdots B_n}$
that can be detected by Theorem \ref{th1} are local unitarily equivalent to $|\phi^+\rangle_{A_1\cdots A_n,B_1\cdots B_n}$.
And if a quantum state is local unitarily equivalent to
$|\phi^+\rangle_{A_1\cdots A_n,B_1\cdots B_n}$, then it must satisfy $\langle \Gamma\rangle_{\rho}=1$ for some local
complementary observables $\{X_k^{A_i}\}_{k=1}^3$, since
$U^{A_1}\otimes \cdots\otimes U^{A_n}\otimes U^{B_1}\otimes \cdots\otimes U^{B_n}|\phi^+\rangle=U^{A_1}(U^{B_1 })^{T}\otimes
\cdots\otimes U^{A_n}(U^{B_n})^{ T}|\phi^+\rangle$
in calculation of the mean values of observables.
Although Theorem \ref{th1} only gives a sufficient condition for the detection of multiqubit teleportation resource, it is experimentally feasible.
Moreover, one only needs to measure different local complementary observables on $A$ part and the observables on $B$ part are fixed.
For example, to detect $|\phi^+\rangle_{A_1\cdots A_n,B_1\cdots B_n}$, one simply chooses $X_k=\sigma_k$, $k=1,2,3$.
Theorem \ref{th1} may also help in the characterization of teleportation witness in multiqubit systems.

\begin{corollary}\label{fully sep det}
If a $2n$-qubit mixed state $\rho$ is fully separable, then $\langle \Gamma \rangle_{\rho}\leq \frac{1}{2^n}$ for all
complementary observables $\{X_k^{A_i}\}_{k=1}^3$ acting on
the qubit $A_i$, $i=1,\cdots,n$.
\end{corollary}

{\sf [Proof].} Due to the linearity of the operator $\Gamma$, we only need to prove $\langle \Gamma \rangle_{\rho}\leq \frac{1}{2^n}$ for all pure fully separable states $|\psi\rangle$.
By proving that the expectation value $\langle|\phi^+\rangle_{A_1\cdots A_n,B_1\cdots B_n}\langle\phi^+|\rangle_{|\psi\rangle\langle\psi|}\leq \frac{1}{2^n}$,
one can prove the corollary directly.
\qed

Corollary \ref{fully sep det} gives a necessary condition for fully separable state. It shows that if the expectation value of $\Gamma$ is
more than $\frac{1}{2^n}$ for some complementary observables, then quantum state is entangled.

\section{Bipartite teleportation witness}

Now we consider the detection of ideal resource for teleportation of high dimensional systems.
Let $H_n$ be an $n$-dimensional complex vector space, with $\{\vert i\rangle\}_{i=0}^{n-1}$ an orthonormal basis.
Let $\rho$ be a density matrix in $H_n\otimes H_n$.
The optimal fidelity of teleportation with $\rho$ as the entangled resource is given by \cite{bennett,M. Horodecki1999-60,alb2002}
\be\label{f}
f_{max}(\rho)=\frac{n F(\rho)}{n+1}+\frac{1}{n+1}.
\ee
$F(\rho)$ is the fully entangled fraction with respect to $\rho$:
\begin{eqnarray}\label{FEF}
F(\rho)=\max_U \langle \psi^+| (U^\dagger \otimes I_n)\, \rho\, (U \otimes I_n) |\psi^+\rangle,
\end{eqnarray}
where $U$ is any $n\times n $ unitary matrix, $I_n$ is the $n\times n$ identity matrix, and $|\psi^+\rangle$
is the maximally entangled state,
$$
|\psi^+\rangle=\frac{1}{\sqrt{n}}\sum_{i=0}^{n-1}|ii\rangle.
$$

A state $\rho$ is a useful resource for teleportation if and only if $F(\rho)>\frac{1}{n}$ \cite{M. Horodecki1999-60}.
If $F(\rho)\leq\frac{1}{n}$, the fidelity (\ref{f}) is no better than separable states. Recently in Ref. \cite{N. Ganguly},
the authors show that the set of states which are not useful
for quantum teleportation, i.e. their fully entangled fractions are no more than $\frac{1}{n}$, is also convex and compact.
Therefore there exist witness operators which detect some entangled states that are useful for teleportation.

A teleportation witness $W$ is a Hermitian operator such that
(i) $Tr(W\rho)\geq0$ for all states $\rho$ that are not useful for quantum teleportation.
(ii) there exists at least one entangled state $\rho$ which is useful for teleportation such that $Tr(W\rho)<0$ \cite{N. Ganguly,S. Adhikari}.
For entangled state $\rho$, if $Tr(W\rho)<0$, then teleportation witness $W$ could detect $\rho$ as a useful resource for teleportation.
Between two teleportation witnesses $W_1$ and $W_2$, $W_1$ is said to be finer than $W_2$ if $W_1$ could detect all the
entangled states which could be detected by $W_2$. A witness is said to be optimal if there exist no other witnesses that are finer than it \cite{M. Lewenstein}.
It is obvious that any teleportation witness is also an entanglement witness.

Since all separable states are positive under partial transpositions (PPT) and there exist PPT entangled states, therefore the set of all separable states
is a proper subset of the set of PPT states. Similarly, since all PPT states are not useful in teleportation \cite{M. Horodecki1999-60} and there exist
nonpositive partial transpositions (NPT) entangled states that are also
not useful in teleportation \cite{P. Badziag}, the set of PPT states is again a proper subset of the set of states that are not useful in the teleportation.
A teleportation witness is a hyperplane that separates a point which is an entangled state that is useful for quantum
teleportation from the convex and compact set of states
that are not useful in teleportation. An optimal teleportation witness is then a hyperplane that is tangent to the set of
states that are not useful in the teleportation (see Fig. 1).
\begin{center}
\begin{figure}[!h]
{\includegraphics{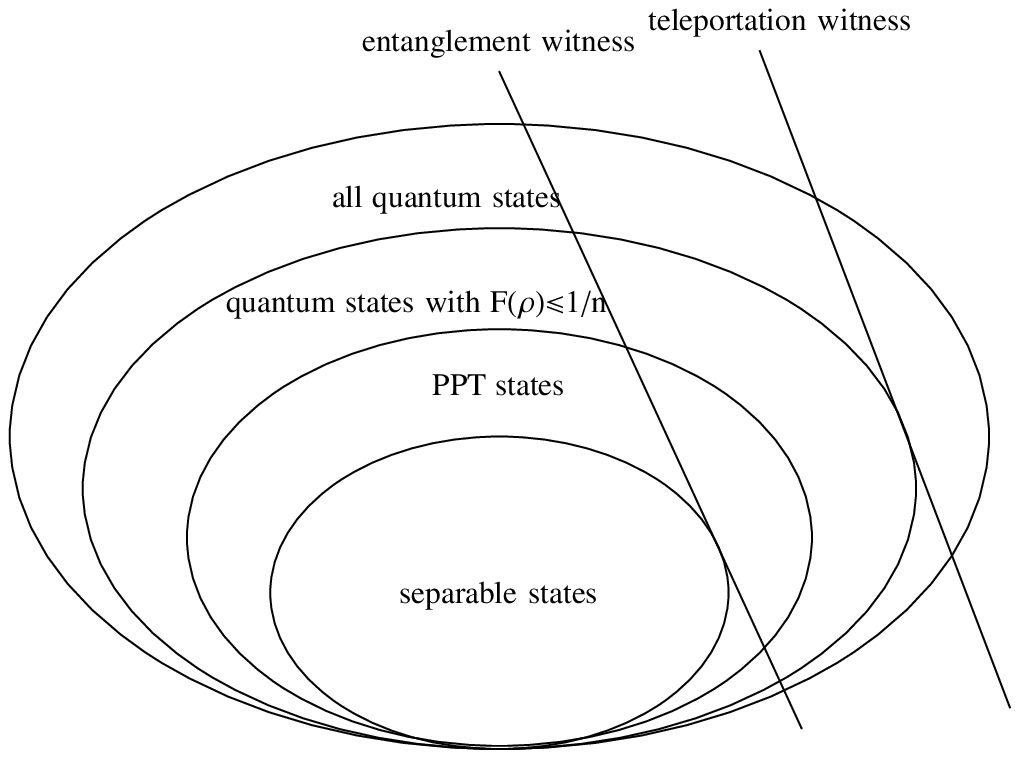}}\caption{}\label{fig1}
\end{figure}
\end{center}

Now we consider the teleportation witness
\begin{eqnarray}
W(I)=\frac{1}{n}I-|\psi^+\rangle\langle\psi^+|.
\end{eqnarray}
$W(I)$ is a teleportation witness in the sense that for arbitrary states $\rho$ which are not useful for quantum teleportation,
i.e, $F(\rho)\leq\frac{1}{n}$, then $Tr(W(I)\rho)\geq0$.
And $Tr(W(I)|\psi^+\rangle\langle\psi^+|)<0$ for the ideal teleportation resource $|\psi^+\rangle\langle\psi^+|$.
Furthermore $W(I)$ is an optimal teleportation witness. Consider the following product vectors:
\begin{eqnarray*}
K_j&=&|jj\rangle,\\
K_{kl}&=&(|k\rangle+|l\rangle)\otimes (|k\rangle+|l\rangle), \\
K_{kl}^\prime&=&(|k\rangle+{\rm i}|l\rangle)\otimes (|k\rangle-{\rm i}|l\rangle),
\end{eqnarray*}
with $j,k,l=0,1,\cdots,n-1$ and $k<l$. These $d^2$ product vectors are linear independent and satisfy $Tr(W(I)K_j)=0$, $Tr(W(I)K_{kl})=0$, $Tr(W(I)K_{kl}^\prime)=0$.
Note that if a set of product vectors $P_W=\{|e,f\rangle: Tr(W|e,f\rangle\langle e,f|)=0\}$ spans the relevant product vector space,
then $W$ is an optimal entanglement witness \cite{M. Lewenstein}. Since an optimal entanglement witness is surly an optimal teleportation witness,
$W(I)$ must be an optimal teleportation witness. In Ref. \cite{S. Adhikari}, a teleportation witness for bipartite system is given and its optimality is proved for two dimensional case and that for high dimensional is left. In fact, one can calculate that $W(I)$ is the exact teleportation witness in Ref. \cite{S. Adhikari}. In this way, the problem proposed in Ref. \cite{S. Adhikari} is solved by proving the optimality of the teleportation witness $W(I)$.

In fact, teleportation witness $W(I)$ is an entanglement witness of the form
$W=\alpha I-|\psi\rangle\langle\psi|$ with $\alpha=\max_{|\phi\rangle=|a\rangle\otimes|b\rangle} |\langle\psi|\phi\rangle|^2$,
where $|\psi\rangle$ is an entangled pure state.
Such entanglement witness $W$ could detect the entanglement of $|\psi\rangle$, and the entanglement of NPT states.
The distance between the entangled state $|\psi\rangle$ and the set of separable states is $\alpha$. Especially, the distance
between the maximally entangled state $|\psi^+\rangle$
and the set of separable states is $\frac{1}{n}$. The distance between $|\psi^+\rangle$ and the set of PPT states is also $\frac{1}{n}$.

Now we define the teleportation witness
\begin{eqnarray}
W(U)=\frac{1}{n}I-U\otimes I|\psi^+\rangle\langle\psi^+|U^\dagger\otimes I.
\end{eqnarray}
One can verify that $W(U)$ is an optimal teleportation witness for arbitrary unitary operator $U$.
Furthermore, the set of teleportation witnesses $\{W(U)\}$ is complete since for arbitrary entangled state $\rho$ that is useful for teleportation,
there exists a unitary operator $U$ such that $Tr(W(U)\rho)<0$.

\section{Conclusion}

We have presented an experimental approach to detect the ideal resource for quantum teleportation for multiqubit systems,
by deriving a condition to detect the local unitary equivalence of the tensor product of Bell states.
Our criterion only involves local measurements on complementary observables in half part of a system.
It also helps in characterizing the teleportation witness for multiqubit
systems \cite{A. Kumar}. For bipartite high dimensional systems, we have analyzed the teleportation witnesses. Moreover, we have solved the open problem presented in Ref. \cite{S. Adhikari}.

\end{document}